\documentclass[referee]{aa}
\usepackage{graphicx}

\begin{document}
\title{Age and metallicity of a globular cluster in the dwarf spheroidal galaxy DDO~78}
\titlerunning{Age and metallicity of a globular cluster in DDO78}
\author{M.E.Sharina \inst{1,3} \and O.K.Sil'chenko \inst{2,4}
\and A.N.Burenkov \inst{1,3}}
\authorrunning{M.E.Sharina et al.}
\institute{Special Astrophysical Observatory, Russian Academy
of Sciences, N.Arkhyz, KChR, 369167, Russia
\and Sternberg Astronomical Institute, Universitetskij Prospect 13, 119992 Moscow, Russia
\and Isaac Newton Institute, Chile, SAO Branch
\and Isaac Newton Institute, Chile, Moscow Branch}
\date{Submitted}

\abstract{
We present the results of moderate resolution spectroscopy for a globular
cluster in the M81 group dwarf spheroidal galaxy DDO~78.
The DDO~78 globular cluster, 4 Milky Way globular clusters,
spectroscopic and radial velocity standards were observed
with the Long-slit spectrograph of the 6-m telescope (SAO RAS, Russia).
Lick spectrophotometric indices were determined in the
bandpasses adopted by Burstein et al. (1984).
We have derived the mean metallicity of the globular cluster in DDO78 to be
[Fe/H] $=-1.6 \pm 0.1$ dex by taking the weighted mean of metallicities
obtained from the strength of several absorption features.
We have estimated an age for the globular cluster of 9-12 Gyr
similar to that found for the Galactic globular cluster NGC~362,
which resembles our cluster by chemical abundance and integrated
spectrophotometric properties.
}
\maketitle

\section{Introduction}
The dominant type of dwarf galaxies in the local Universe
is the dwarf spheroidal galaxies (DSphs). These systems are characterized by
low luminosities ($M_\mathrm{V} > -14$),
low stellar density ($\mu_\mathrm{v}(0) \ga 22 ^m/\sq\arcsec$),
small spatial extent (core radii $<$ a few
hundred parsecs), HI masses
$M_\mathrm{HI} \la 10^5 M_{\sun}$, lack of current star formation (Grebel 2000, Stetson et al. 1998).
 New deep C-M diagrams and spectroscopy of individual red giants in several
Local Group dSphs have shown us that, in spite of their current quiescent
appearance, most of the systems have had surprisingly complicated
evolutionary histories (Harbeck et al. 2002).
 It would be interesting
to compare the chemical and star formation histories for DSphs in the Local
group and in the other nearby groups.  It is a difficult task, because
the observational magnitude limit does not
allow us to observe the HB stars in the galaxies at a distance of $>$ 3 Mpc.
Globular clusters (GCs) to be found in some DSphs are ideal probes
of chemical histories and evolution of their host galaxies.
They are bright enough to be observed at large distances.
If the correlations between key spectral indices, metallicities
and ages are common for GCs in all types of galaxies,
we can determine the age for the given particular GC by
using the evolutionary population synthesis models
(e. g. Worthey 1994) and
compare our results with the data
available in literature on the Milky Way globular clusters.

The group of galaxies around the bright spiral M81 is one of the
nearest prominent groups in the vicinity of the Local Group.
It is situated at a distance of $\sim 3.7$ Mpc
and consists of at least 12 DSphs and 12 dwarf irregular galaxies (Karachentsev et al. 2002).
DDO~78 is one of the faintest DSphs of the group ($M_\mathrm{V}=-12.83$).
According to Karachentseva et al. (1987), DDO~78 has a very flat surface
brightness profile with $\mu_\mathrm{B(0)} = 25.1^m/\sq\arcsec$, $r_\mathrm{ef}=32\arcsec$, and $B(t)=15.8$.
It was not
detected in H{\sc I} by Fisher \& Tully (1981) and van Driel et al. (1998).
An accurate distance modulus of DDO~78 is $(m-M)_0=27.85 \pm 0.15$
according to Karachentsev et al. (2000).
The latter authors found globular cluster candidates in five dwarf spheroidal
galaxies of the M81 group and measured their basic photometric parameters.
The brightest globular cluster candidate in DSph galaxy DDO~78
has the integrated apparent magnitude $V_\mathrm{t}=19.45$,
the integrated color after correction for Galactic reddening $(V-I)_{0}=1.07$,
the angular half-light radius $R(0.5L)=0.3 \arcsec$,
the central surface brightness
$\mu_v(0)=18.0 ^m/\sq\arcsec$,
the linear projected separation of the globular cluster from the galaxy
center 0.26 kpc.
Sharina et al. (2001) measured a heliocentric radial velocity of the
globular cluster candidate in DDO~78
to be  $55\ \pm 10\  \mathrm {km}\ \mathrm{s}^{-1}$ by cross-correlation
with template stars and established the object as a {\it bona fide}
member of the galaxy.
The present work continues our study of the dwarf spheroidal galaxies in the M81 group.
The spectra of the globular cluster in DDO~78 (Fig.1) are of rather high
signal-to-noise ratio  and are suitable for the quantitative
spectrophotometric study.

\section{Observations and data reduction}
The observations were performed with the Long-slit spectrograph UAGS
(Afanasiev et al. 1995) at the prime focus of the 6-m telescope (SAO RAS, Russia)
during 4 nights in January 2001 at a seeing of $\sim 1 \arcsec$ (see Table 1
for details).  An additional observation of stars from the Lick/IDS sample
of Worthey et al. (1994) was performed in April 2002.
 The long-slit $130\arcsec$ spectra were obtained with a
CCD-detector having 1024x1024 pixels with 24x24 $ \mu $m
pixel size. For all observations we used a grating of
651 grooves/mm with a corresponding dispersion of 2.4 \AA/pixel and a spectral
resolution of 7-9 \AA \hspace{0.5mm}.
The slit positions were chosen to cross the star cluster.
The wavelength range was 4500-6900\AA\AA. In all the cases
the slit width was 2\arcsec \hspace{0.5mm}. The scale along the slit
was 0.41\arcsec/pixel. The reference spectra of Ar-Ne-He lamp
were exposed before and after each observation to provide
wavelength calibration.
For velocity calibration we have obtained long-slit spectra of seven
radial-velocity standard stars (Barbier-Brossat M. \& Figon P. 2000)
(see Table 1).
 The spectrophotometric standard stars BD28 4211, HZ44 and Feige34 (Bohlin 1996)
were observed for flux calibration.

 The data reduction was performed using the LONG package
in MIDAS. The subsequent data analysis was also carried out in MIDAS.
 The primary data reduction included cosmic-ray removal,
bias subtraction and flat-field correction. After wavelength
calibration and sky subtraction, the spectra were corrected for atmospheric
extinction and flux-calibrated.
Then rows of every linearized two-dimensional spectrum were summed
in the spatial direction to yield a final one-dimensional spectrum.
All individual exposures of the same object observed at the same night
were then co-added to increase the S/N ratio.

 In order to obtain integrated spectra for the Milky Way clusters (Fig.2),
we averaged the spectra of different zones and individual stars in the core
of each cluster. The sizes of the integrated apertures along the slit,
equatorial coordinates of the GC centers and positional angles
of the slit are listed in Table~2.
All the spectra of the given particular GC were recorded at the same slit
position across the center of the object.

Radial velocities of the globular clusters
have been obtained by cross-correlation with template stars
and were used for a subsequent abundance analysis.
A comparison of the radial
velocities measured by us with the corresponding values from Harris (1996)
(Table~3, this paper) shows a good agreement.

\section{Measurement of spectral indices and errors}
Absorption-line indices were measured for each object according
to the index definitions reported by Burstein et al. (1984).
Each line index is a measurement of the flux, contained in the wavelength
region centered on the feature relative to that contained in the red and
blue "continuum" regions close to the features (Faber et al. 1977).
We computed the index $I$ as:
$$ I = -2.5 \log \frac {F_\mathrm{I}} {F_\mathrm{C1} + k(F_\mathrm{C2} - F_\mathrm{C1})},$$
where $F_\mathrm{I}$ is the mean flux in the feature, and $F_\mathrm{C1}$,
$F_\mathrm{C2}$ are the mean fluxes in the continuum  bandpasses. $k$ is
a constant defined by interpolation between the central wavelengths
in the blue and red bandpasses with respect to the central wavelength of
the bandpass centered on the feature:
 $$ k= \frac {(\lambda_{F_\mathrm{I_2}} - \lambda_{F_\mathrm{C1_2}}) +
(\lambda_{F_\mathrm{I_1}} - \lambda_\mathrm{F_\mathrm{C1_\mathrm{1}}})}
{(\lambda_\mathrm{F_\mathrm{C2_\mathrm{2}}} - \lambda_\mathrm{F_\mathrm{C1_\mathrm{2}}})
 + (\lambda_{F_\mathrm{C2_\mathrm{1}}} - \lambda_\mathrm{F_\mathrm{C1_\mathrm{1}}})}.$$
It is possible to convert the index values from the magnitudes to pseudo-equivalent widths and vice
versa by the formulae:
 $$ EW(I)=( \lambda_\mathrm{2} - \lambda_\mathrm{1})(1-10^{-\frac{I}{2.5}})$$
 $$ I(EW)=-2.5 \log (1- \frac{EW} {\lambda_\mathrm{2} - \lambda_\mathrm{1}}).$$

We have derived the errors of the measurements of the line indices
according to equations of Cardiel et al. (1998).
The most common sources of systematic errors in the measurement of
line-strength indices are:
flux calibration effects, spectral resolution, sky subtraction uncertainties,
scatter light effects, wavelength calibration and radial velocity errors,
seeing and focus corrections, response deviations from linearity
and random Poisson noise. We used a correlation between the measured absolute
index error and the mean (S/N-ratio)/\AA \hspace{0.5mm} studied by Cardiel
et al. (1998) to derive the error of our line-strength index
measurements.
We computed the mean signal-to-noise ratio in each bandpass via the sum of the
individual S/N ratios in each pixel:
$$ SN (\mathrm \AA) = \frac {1} {N \sqrt{ \theta }} \sum_{i=1}^{N} \frac {S( \lambda_\mathrm{i})}{ \sigma ( \lambda_\mathrm{i})},$$
where $ \theta$ is the dispersion (in \AA /pixel).

Two spectra of the globular cluster in DDO~78 were obtained
during two nights under similar observational conditions (Table 1, Fig.1.).
The final index values for the GC in DDO~78 (Table 5) are
a weighted average of the indices from the two spectra
using $\frac{1}{\sigma_\mathrm{i}^2}$ as a weight.
The final index photon errors, $\sigma_\mathrm{I}$, are given
by the reduced photon error:
$$ \sigma_\mathrm{I} = (\sum_{i=1}^{N}\frac{1}{\sigma_\mathrm{i}^2})^{-1/2}.$$
To check the calibration of our index measurements into the standard Lick
system, we observed 4 stars from the list of Worthey et al. (1994).
A comparison of the measured long-slit indices with those from
Worthey et al. (1994) is given in Table 4 together with the measured
indices of the twilight sky. The Lick indices of the twilight sky are known
for us from the extensive observations with the Multi-Pupil Fiber
Spectrograph of the 6~m telescope that are well calibrated into the
standard Lick system through the careful monitoring of the standard stars
(the last and most complete calibration of the MPFS index system
into the standard Lick one can be found in Afanasiev \& Sil'chenko 2002).
One can see that our index system coincides with the standard Lick one
within the errors of measurements.
Besides the stars, we observed also
the cores of 4 galactic clusters NGC~5272, NGC~2419, NGC~4147 and Pal~1
to compare them with the GC in DDO~78.
The candidates for observations were chosen to conform to an expected
metallicity of the globular cluster in DDO78.
The measured indices in the Lick system for the globular
clusters are listed in Table 5.

To investigate the properties of the GC in DDO78 we supplemented our data
by observations of Cohen et al. (1998), Covino et al. (1995),
Burstein et al. (1984), and Brodie and Huchra (1990) for
galactic globular clusters, Huchra et al. (1996) for the Fornax dSph
globular clusters and Schroder et al. (2002) for the M81 globular clusters.
All observations were made in the Lick system.
The data on $M_\mathrm{V}$, $V$ and $I$ for the galactic GCs
were taken from the catalog of Harris (1996).

\section{Globular cluster metallicity and age}
Metallicities for the observed GCs were calculated using the linear
least-square
fits of the indices versus metallicity for metallicity-sensitive line
indices made by Covino et al. (1995).
We obtained the mean metallicity of the cluster in DDO~78 and
the mean metallicities of Galactic GCs (Table 6)
by taking the error-weighted
average of the metallicities predicted by the strength of five
metallicity-sensitive absorption-line indices:
Mgb, Fe5270, Fe5335, Mg$_\mathrm{1}$ and Mg$_\mathrm{2}$.
 It should be noticed, that
Mg$_\mathrm{1}$ and Mg$_\mathrm{2}$ each was assigned half weight,
because these indices are tightly correlated (Burstein et al. 1984).
A comparison of the indices measured by us for the core of
the Milky Way GC NGC~5272$=$M~3 with the corresponding values from
Burstein et al. (1984) shows a good agreement.
 The estimates of [Fe/H] derived by us for Galactic GCs NGC~2419,
NGC~4147, and M~3 agree also rather well with the catalogue values:
in the last column of Table~6 the estimates from Zinn \&\ West are given for
every cluster. As for Pal~1, whose metallicity given in Table~6 has been
found by Rosenberg et al. (1998a), this cluster is known to be much younger
than the bulk of other Galactic globular clusters ($T(Pal1)=6-8$ Gyr,
Rosenberg et al. 1998b); so the calibration
of metallicity-sensitive indices made by using the old globular clusters
is inapplicable to it.
The value [Fe/H] $=-1.6 \pm 0.1$~dex calculated in the present work for
the GC in DDO78 does not contradict the mean metallicity of the galaxy
itself derived from the red giant branch morphology by Karachentsev et al.
(2000), [Fe/H] $=-1.6 \pm 0.3$ dex.

Fig. 3a shows the index $\langle \mbox{Fe} \rangle = (Fe5270 + Fe5335)/2$ as a function
of the index Mgb. It is evident that the GC in DDO78 follows well Worthey
models and is similar to the galactic globular clusters on
this diagram. A comparison of the metallicity-sensitive indices with the
corresponding values for the
Galactic GC with [Fe/H]~$\sim -1.6$~dex shows a good agreement.
In Fig. 3b, 3c, and 3d we plot the strengths of
Mg$_\mathrm{2}$, Mgb, and $\langle \mbox{Fe} \rangle$ as a function of
[Fe/H].
The low-metallicity Galactic and M81 GCs and GC in DDO78 occupy generally
the same positions in these Figures.

We continue our analysis by plotting various indices
as a function of $(V-I)$ color corrected for extinction, adopting the photometry of
Karachentsev et al. (2000) for the globular cluster in DDO78.
Fig. 4a, 4b, 4c, and 4d show $(V-I)_\mathrm{0}$ color versus Mgb,
Mg$_\mathrm{2}$, $\langle \mbox{Fe} \rangle$ and [Fe/H].
It is clear that the GC in DDO78 falls rather into a group of
low-metallicity Galactic GGs.
However, its $(V-I)_\mathrm{0}$ color seems to be too red.

 $(V-I)_\mathrm{0}$ colors for M81 GCs were computed from
$(B-V)$ colors derived by Schroder et al. (2002) and corrected for foreground
reddening and for the mean value of internal reddening in M81 computed by the latter authors.
We used the relationship between intrinsic $(B-V)$ and $(V-I)$ colors
for Milky Way GCs with the integrated spectral type F9/G0-G2 from Reed et al. (1988).

To obtain the age of the GC in DDO~78 we plot the age-sensitive index
H$\beta $ against Mgb (Fig. 5a),
Mg$_\mathrm{2}$ (Fig. 5b), $\langle \mbox{Fe} \rangle$ (Fig. 5c),
and [Fe/H] (Fig. 5d).
The signs are the same as in the previous figures.

Fig. 5 demonstrates that the GC in DDO~78 has a too low
H$\beta$ for its weak metal absorption lines: it falls below the oldest
model sequence, $T=17$ Gyr. However, it is not a single outlier: the whole
group of the Galactic globular clusters with red horizontal branches,
mostly of intermediate metallicity, are also well below the 17-Gyr model
sequence of Worthey (1994). The problem is not our finding: Covino et al.
(1995), comparing their index measurements for Galactic globular clusters
with the models of Buzzoni et al. (1992, 1994), which included an EMPIRICAL
RED horizontal branch, have also found the same shift. They have suggested
that this discrepancy results from the solar [$\alpha$/Fe] ratio of the
models: if one takes $\alpha$-enhanced isochrones, the isochrone turn-off
points would be redder (cooler), and the corresponding H$\beta$
equivalent widths would be lower. Since the Galactic globular clusters
are known to have [O/Fe]$=+0.3 \div +0.5$, their suggestion seemed to be
the solution. However, during the last years, there were numerous attempts
to improve model isochrones and evolutionary synthesis approach to the
globular clusters, and none of them gives anything to correct the
discrepancy of Fig.5. Salaris and Weiss (1998) have
demonstrated their new $\alpha$-enhanced isochrones, from which one can
see that for the intermediate metallicity, [Fe/H]~$=-1.35$~dex, the
$\alpha$-element enhancement affects the slope of the giant branch
on the C-M diagram, but leaves the main-sequence turn-off position
unaffected. Maraston (1998) (see also Maraston et al. 2001) has revised
the H$\beta$ estimates by evolutionary synthesis with new isochrones and
the fuel-consumption theorem application to the horizontal branch
treatment, but her H$\beta$ indices are even higher than those of
Worthey (1994) and Buzzoni et al. (1992, 1994).

So, a conclusion must be made that up to now state-of-art models cannot
explain some intermediate-metallicity globular clusters, namely, almost
all the clusters with red horizontal branches. Under such
circumstances the only reasonable way to determine the age of the GC in DDO~78
is to compare it with any particular Galactic globular cluster for which
a detailed C-M diagram is available. The nearest neighbor of the GC in
DDO~78 in Fig.5 is a well-studied Galactic globular cluster
NGC~362. While the metallicities of the most globular clusters with
red horizontal branches are higher than [Fe/H]$=-1$~dex, the metallicity
of NGC~362 is only slightly higher than that of the GC in DDO~78,
namely, its [Fe/H]~$=-1.33 \pm 0.01$~dex (Shetrone \& Keane 2000), and it is
a member of the classical second-parameter cluster pair NGC~288/NGC~362.
These two clusters have the same metallicities, but quite different
morphologies of the horizontal branches, the horizontal branch of NGC~362
being a red one. After a long discussion about the nature of the second
parameter determining a morphology of the horizontal branches, a common
view is established that it may be an age; in particular, a careful
examination of the C-M diagrams of NGC~288 and NGC~362 by Bellazzini et al.
(2001) and Catelan et al. (2001) has proved that the latter cluster is
younger by 1.5--2 Gyr than the former. As for the estimates of the absolute
age of NGC~362, they range from 8.7 Gyr (Salaris \&\ Weiss 1998) and
9.9 Gyr (Carretta et al. 2000) to 11--12 Gyr (Vandenberg 2000). Since
the differences of H$\beta$, Mg$_2$, Mgb, and $\langle \mbox{Fe} \rangle$
indices between NGC~362 and GC in DDO~78 are practically within the
observational errors, we can suggest that the age of the GC in DDO~78
is also about 9--12 Gyr.

\section{Discussion}
 According to Burgarella et al. (2001) the mean metallicity
of metal-poor globular cluster systems is weakly dependent on the
host galaxy's properties and is almost "universal" at [Fe/H]~$\sim-1.4 \pm 0.3$~dex.
 The metallicity of the GC in DDO78 [Fe/H]~$=-1.6 \pm 0.1$~dex
agrees well with this value.
There are two DSph companions of the Milky Way which contain globular clusters:
Fornax and Sagittarius DSph galaxies.
Fornax DSph has the absolute integral magnitude $M_\mathrm{V} = -13.7$, and
the distance from the Milky Way 140 kpc (van den Bergh 2000).
These properties are comparable to those of DDO~78, which is located at the
distance of 223 kpc from M81 (Karachentsev et al. 2002).
Five Fornax DSph globular clusters have
the mean metallicity [Fe/H]~$=-2.04$ dex, essentially the
same ages ( $\mid \delta t \mid < 1$ Gyr) and are coeval with the
old, metal-poor clusters of our Galaxy (Buonanno et al. 1998)
Unfortunately, there are no measured integrated absorption indices for
all Fornax GCs in the literature.
Fornax DSph GCs 3 and 5 show absorption feature strengths (Huchra et al. 1996)
unlike those for the GC in DDO78.
Our cluster seems to be more metal-rich than the Fornax DSph GCs.
We found only one Galactic globular cluster with measured integral
absorption indices, NGC~362,
which resembles the GC in DDO78 by all properties within the errors.
NGC~362 has the galactocentric radius $R_{gc}=9.2$ kpc, the integral absolute visual
magnitude $M_\mathrm{V} =-8.4$ and resembles by its properties
some intermediate-metallicity clusters with $R_{gc} > 8$ kpc, Pal~12, NGC~1851,
NGC~1261, NGC~2808 (Rosenberg et al. 1999). These clusters are
younger by $ \sim 2$ Gyr than the metal-poor halo GCs and are associated with
so-called `streams' that may be relics of ancient Milky Way satellites
which had masses typical of a dwarf galaxy.

The M81 group of galaxies has a similar to the LG structure (Karachentsev et al. 2000).
The subgroups around the Sb type galaxies, the Milky Way and M31,
have a spatial separation of $\sim 1$ Mpc and approach to each other at a velocity
of $\sim 130$ $ \mathrm {km}\ \mathrm{s}^{-1}$.
Such a situation resembles the M81/NGC~2403 complex.
But there are some differences between the two groups.
The core members of the M81 group: M81, M82, NGC~3077 and NGC~2976
are known to be closely interacting from the
aperture synthesis maps of the 21 cm H{\sc I} emission (Yun et al. 1994, Boyce et al. 2001).
The dwarf spheroidal galaxies are distributed around M81 asymmetrically
(Karachentsev et al. 2001). With respect to the group centroid
(located between M81 and M82) all DSphs are concentrated in one
quadrant.
Jonson et al. (1997) reported about the discovery of the high-excitation
H{\sc II} region in  K61, the brightest DSph galaxy of the M81 group
and the closest companion to M81.
They found the metallicity of H{\sc II} region, Z,
between 0.001 and 0.008, and the age between 2 and 5.2 Myr.
The properties of DSph galaxies appear to be correlated with the galaxy mass
and with environment.
Does the tidal interaction between the brightest M81 group
galaxies influence the distribution and star formation histories
of DSphs? Accurate numerical simulations are needed.

\acknowledgements{
We thank the anonymous referee for helpful comments.

The 6~m telescope of the Special Astrophysical Observatory (SAO)
of the Russian Academy of Sciences (RAS)
is operated under the financial support
of the Science Department of Russia (registration number 01-43).}

\begin{table}[hbt]
\caption{Observational log}
\scriptsize
\begin{tabular}{lclc} \\ \hline
Object            & Date             & Exposure & Seeing         \\ \hline
Globular cluster  &   18.01.2001 &  2 x 1200 s  & 1.$\arcmin$3      \\
in DDO78          & 18.01.2001   & 2 x 2400     &      \\
		  &   19.01.2001 &  8 x 1200    &  1.3 \\
\\
NGC2419           & 19.01.2001   &  2 x 600     & 1.3  \\
\\
NGC4147           & 19.01.2001   & 2 x 600      & 1.3  \\
\\
Pal1              & 22.01.2001   & 3 x 900      & 2.2  \\
\\
NGC5272           & 22.01.2001   & 3 x 600      & 2.2  \\
\\
BF10078 (F8 V)   & 18.01.2001  &  3 x 10  & 1.3    \\
BF11123 (G8 V)   & 18.01.2001  &  3 x 30  & 1.3    \\
BF13987 (F8 VI)  & 18.01.2001  &  3 x 60  & 1.3    \\
BF18804 (G9 V)   & 19.01.2001  &  2 x 20  & 1.3    \\
BF18757 (G9 V)   & 19.01.2001  &  2 x 10  & 1.3    \\
BF15344 (F8 V)   & 22.01.2001  &  2 x 20  & 2.2     \\
BF16879 (G8 IV)  & 22.01.2001  &  2 x 20  & 2.2    \\
\\
HD073593 (G8IV)   & 18.04.2002  &  2 x 5   & 4.0  \\
HD091739 (G0)     & 18.04.2002  &  2 x 40  & 4.0  \\
HD152792 (G0V)    & 18.04.2002  &  2, 5    & 4.0  \\
HD142373 (F8Ve)   & 18.04.2002  &  2 x 1   & 4.0  \\
\\
BD28 4211      &  18.01.2001 &  2 x 90     & 1.3  \\
	       & 19.01.2001 &  2 x 150     & 1.3  \\
Feige 34       &  22.01.2001 & 2 x 120     & 2.2  \\
HZ 44          &  18.04.2002  & 240, 120   & 4.0  \\
\hline
\end{tabular}
\end{table}
\begin{table}[hbt]
\caption{Additional observational data on the Milky Way globular clusters}
\scriptsize
\begin{tabular}{lccl} \\ \hline
Object          & RA  (2000.0)  Dec          & PA of        & Aperture  \\
		&                            & the slit     & along the slit    \\  \hline
NGC2419         & 07 38 08.22  +38 52 56.42   & 152.$^{o}$6   &   14$\arcsec$       \\
\\
NGC4147         & 12 10 06.66  +18 32 34.31   & 102.6       &   20                \\
\\
Pal1            & 03 33 20.85  +79 34 57.33   & 151.2       &   7                  \\
\\
NGC5272         & 13 42 11.56  +28 22 32.60   & 51.8        &   60                 \\
\hline
\end{tabular}
\end{table}
\begin{table}
\caption{Measured heliocentric radial velocities of the globular clusters in comparison with corresponding data from Harris (1996).}
\begin{tabular}{lcc} \\ \hline
Object               & $V_\mathrm{h}$ (our), $ \mathrm {km}\ \mathrm{s}^{-1}$ & $V_\mathrm{h}$ (Harris), $ \mathrm {km}\ \mathrm{s}^{-1}$ \\
\hline
GG in DDO78          & $55 \pm 10$     & ---  \\
NGC4147              & $189 \pm 11 $   & 183.2  \\
NGC2419              & $ -4.4 \pm 23 $ & -20.0  \\
NGC5272              & $ -130 \pm 15 $ & -148.5  \\
Pal 1                & $ -60 \pm 25 $    & -82.8   \\
\hline
\end{tabular}
\end{table}

\begin{table*}[hbt]
\caption{ Calibration of our index measurements into the standard Lick system.
Top: A comparison of the Lick indices measured with the Long-slit spectrograph
UAGS (6~m telescope) with those from Worthey et al. (1994) for the observed stars.
Bottom: A comparison of the twilight sky Lick indices measured
with the Multi-Pupil Fiber Spectrograph of the 6~m telescope
with those measured with the UAGS (see text for details).
The values of atomic indices are given in \AA.
The values of molecular indices are given in magnitudes.}
\begin{tabular}{lrrrrrr} \\ \hline \hline
Object      & H$ \beta$         & Mgb            & Fe 5270       & Fe5335        &  Mg$_\mathrm{1}$& Mg$_\mathrm{2}$ \\
\hline \hline
HD073593 (Lick/IDS)& 1.38$\pm$ 0.22 & 3.08$\pm$0.23  & 2.96$\pm$0.28 & 2.16$\pm$0.26  & 0.072$\pm$0.007 & 0.183$\pm$0.008 \\
HD073593 (UAGS)    & 1.10$\pm$0.13  & 2.75$\pm$0.30  & 2.99$\pm$0.43 & 2.22$\pm$0.33  & 0.064$\pm$0.018 & 0.138$\pm$0.021 \\
\\
HD091739 (Lick/IDS)& 3.75$\pm$0.22 & 1.09$\pm$0.23  & 1.35$\pm$0.28 & 1.25$\pm$0.26   & 0.013$\pm$0.007 & 0.070$\pm$0.008 \\
HD091739 (UAGS)    & 3.58$\pm$0.11 & 1.13$\pm$0.12  & 1.20$\pm$0.13 & 0.86$\pm$0.20   & -0.014$\pm$0.007 & 0.033$\pm$0.008 \\
\\
HD152792 (Lick/IDS)& 2.02$\pm$0.22 & 1.82$\pm$0.23  & 0.92$\pm$0.28 & 1.10$\pm$0.26   & ---       & 0.084$\pm$0.008 \\
HD152792 (UAGS)    & 2.17$\pm$0.19 & 2.03$\pm$0.13  & 1.57$\pm$0.21 & 1.07$\pm$0.19   & 0.007$\pm$0.011 & 0.083$\pm$0.013 \\
\\
HD142373 (Lick/IDS)& 2.26$\pm$0.22 & 2.01$\pm$0.23  & 0.71$\pm$0.28 & 0.96$\pm$0.26   & 0.007$\pm$0.007 & 0.078$\pm$0.008 \\
HD142373 (UAGS)    & 2.37$\pm$0.20 & 1.76$\pm$0.12  & 1.21$\pm$0.20 & 0.92$\pm$0.26   & 0.003$\pm$0.010 & 0.062$\pm$0.011 \\
\hline
Twilight sky (MPFS) & 1.86$\pm$0.03 & 2.59$\pm$0.05 & 2.04$\pm$0.06 & 1.59$\pm$0.11    & 0.002$\pm$0.004 & 0.066$\pm$0.002 \\
Twilight sky (UAGS) & 1.97$\pm$0.04 & 2.52$\pm$0.02 & 2.13$\pm$0.07 & 1.76$\pm$0.12    & -0.033$\pm$0.010 & 0.012$\pm$0.006 \\
\hline \hline
\end{tabular}
\end{table*}

\begin{table*}[hbt]
\caption{Calibrated indices and errors for the observed globular clusters.
The values of atomic indices are given in \AA.
The values of molecular indices are given in magnitudes.}
\begin{tabular}{lrrrrrr} \\ \hline
Object      & H$ \beta$  & Mgb     & Fe5270     & Fe5335     & Mg$_\mathrm{1}$     & Mg$_\mathrm{2}$  \\
\hline
GC in DDO78 & 1.36       & 0.80       & 0.87       & 0.55       & 0.023      & 0.037      \\
	    & $\pm$0.23  & $\pm$0.27  & $\pm$0.37  & $\pm$0.40  & $\pm$0.015 & $\pm$0.018  \\
NGC2419     & 2.16       & 0.37       & 0.92       & 0.68       & -0.024     & -0.004      \\
	    & $\pm$0.54  & $\pm$0.49  & $\pm$0.45  & $\pm$0.831 & $\pm$0.025 & $\pm$0.030  \\
NGC4147     & 2.19       & 0.66       & 0.78       & 0.67       & 0.005      & 0.035       \\
	    & $\pm$0.12  & $\pm$0.15  & $\pm$0.19  & $\pm$0.098 & $\pm$0.006 & $\pm$0.008  \\
Pal 1       & 2.17       & 1.79       & 1.56       & 1.40       & 0.029      & 0.097       \\
	    & $\pm$0.43  & $\pm$0.43  & $\pm$0.52  & $\pm$0.632 & $\pm$0.020 & $\pm$0.024  \\
NGC5272     & 2.88       & 1.07       & 1.19       & 0.96       & 0.007      & 0.059       \\
	    & $\pm$0.08  & $\pm$0.18  & $\pm$0.22  & $\pm$0.16  & $\pm$0.008 & $\pm$0.009  \\
\hline
\end{tabular}
\end{table*}
\begin{table*}[hbt]
\caption{Metallicity predicted by each spectral feature and the error
weighted metallicity of each cluster.}
\begin{tabular}{lrrrrrrr} \\ \hline
Object       & Mgb     & Fe5270     & Fe5335     & Mg$_\mathrm{1}$     & Mg$_\mathrm{2}$    & [Fe/H]$_{our}$ & [Fe/H]$_{ZW}$ \\
\hline
GC in DDO78  & -1.60      & -1.78      & -1.88      & -1.35      & -1.79      & -1.6     & --         \\
	     & $\pm$0.13  & $\pm$0.29  & $\pm$0.41  & $\pm$0.30  & $\pm$0.18  & $\pm$0.1  &            \\
NGC2419      & -1.81      & -1.73      & -1.75      & -2.30      & -2.21      & -2.08     & -2.10      \\
	     & $\pm$0.55  & $\pm$0.83  & $\pm$1.94  & $\pm$0.50  & $\pm$0.30  & $\pm$0.24 & $\pm$0.30  \\
NGC4147      & -1.67      & -1.85      & -1.76      & -1.71      & -1.82      & -1.75     & -1.80      \\
	     & $\pm$0.17  & $\pm$0.34  & $\pm$0.23  & $\pm$0.13  & $\pm$0.08  & $\pm$0.07 & $\pm$ 0.26 \\
Pal 1        & -1.09      & -1.21      & -1.00      & -1.24      & -1.19      & -1.17     & -0.6       \\
	     & $\pm$0.49  & $\pm$0.94  & $\pm$1.47  & $\pm$0.40 & $\pm$0.24  & $\pm$0.21 & $\pm$ 0.2  \\
NGC5272      & -1.46      & -1.51      & -1.46      & -1.67      & -1.57      & -1.56     & -1.66      \\
	     & $\pm$0.20  & $\pm$0.40  & $\pm$0.38  & $\pm$0.15  & $\pm$0.09  & $\pm$0.08 & $\pm$ 0.06 \\
\hline
\end{tabular}
\end{table*}

\clearpage
\setcounter{figure}{0}
\begin{figure}[hbt]
\begin{tabular}{p{0.9\textwidth}}
\centerline{\includegraphics[width=10.0cm,angle=-90]{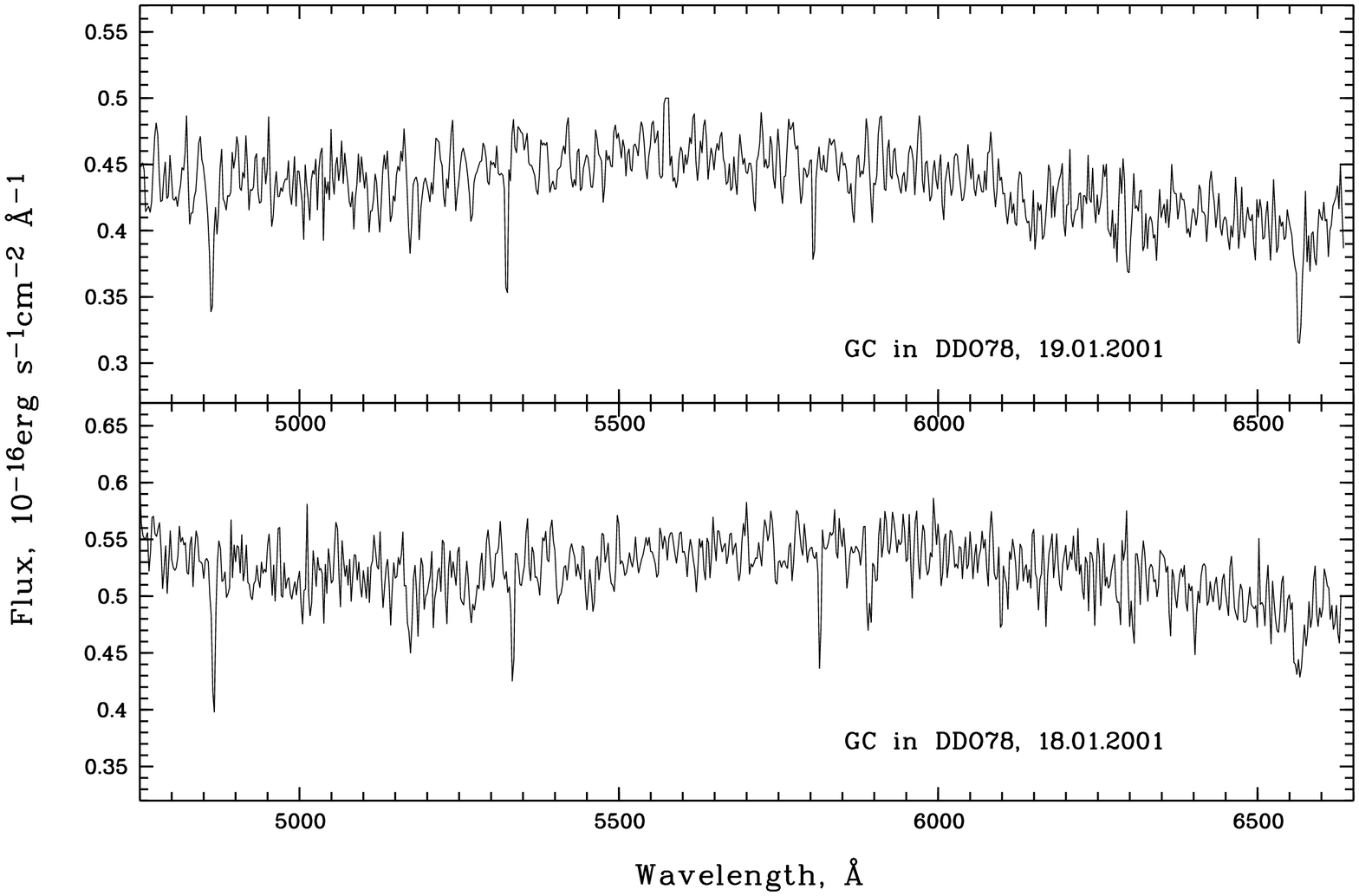}} \\
\caption{Two spectra of the globular cluster in DDO78 obtained on
18.01.2001 (top) and on 19.01.2001 (bottom).} \\
\setcounter{figure}{1}
\centerline{\includegraphics[width=10.0cm,angle=-90]{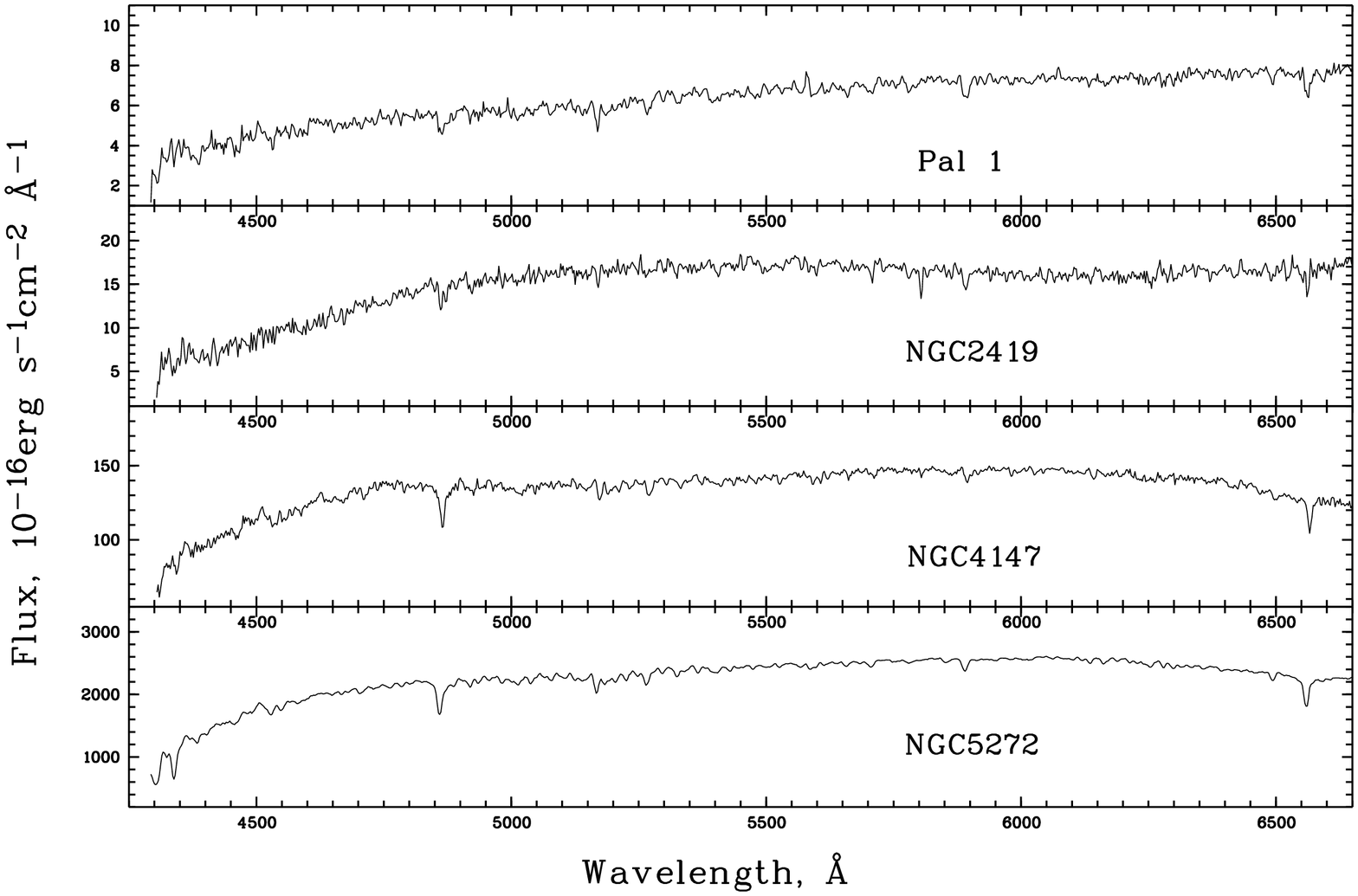}} \\
\caption{Spectra of Galactic globular clusters NGC5272, NGC2419, NGC4147 and Pal1 observed by us.} \\
\end{tabular}
\end{figure}

\setcounter{figure}{2}
\begin{figure*}
\hspace{-1.0cm}
\begin{tabular}{p{0.90\textwidth}}
\includegraphics[width=14.7cm,clip]{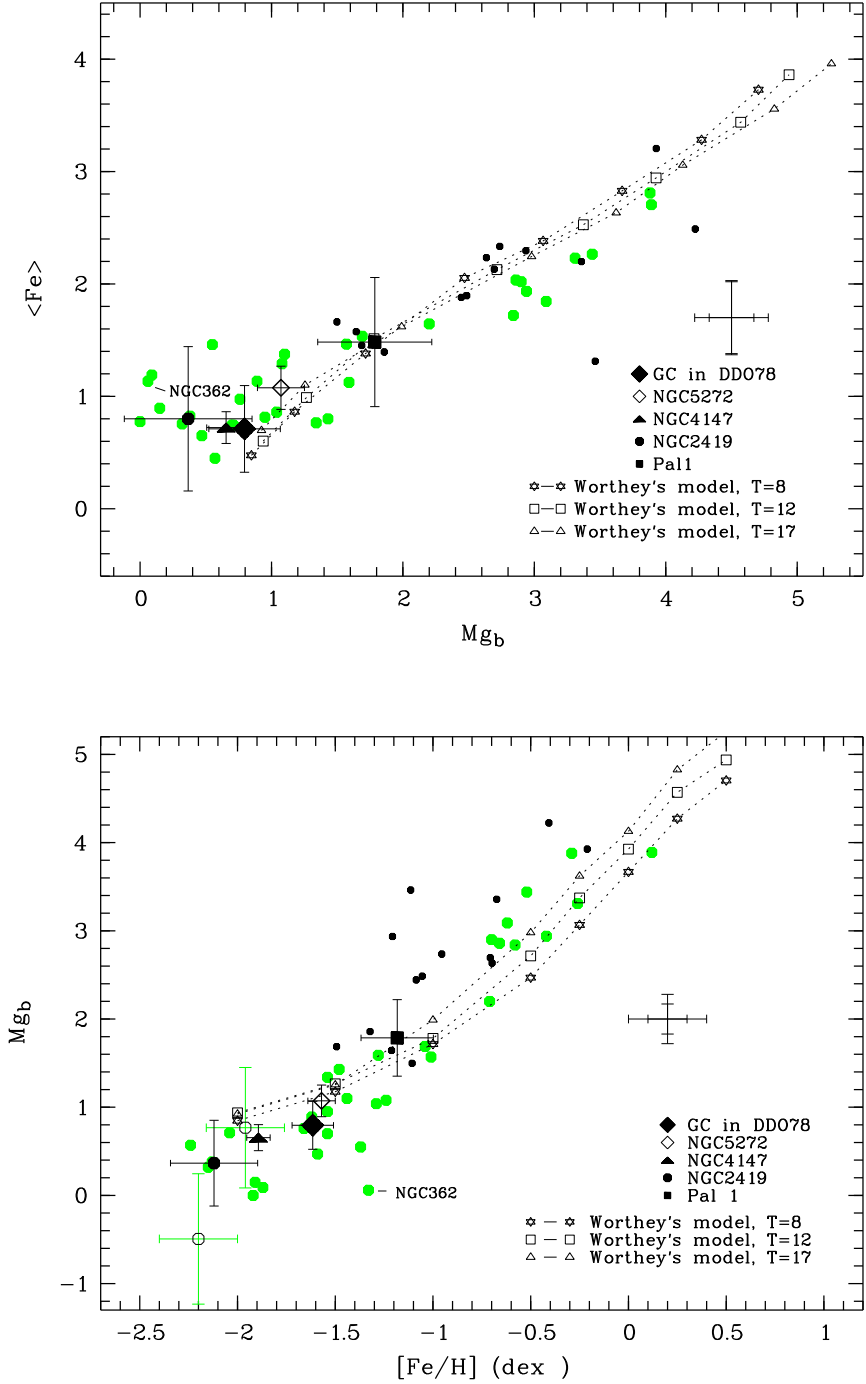} \\
\end{tabular}
\caption{ $\langle \mbox{Fe} \rangle=(Fe5270 + Fe5335)/2$
plotted against Mgb index (Fig.3a); also we show
[Fe/H] correlations with the metallicity-sensitive indices Mg$_2$ (Fig.3b),
 Mgb (Fig.3c), and $\langle \mbox{Fe} \rangle$ (Fig.3d)
for the observed globular clusters (signs marked on the plot), for
Milky Way (grey dots), M81 (dark small dots) and
Fornax DSph globular clusters (open circles).
Typical errors for the M81 clusters are shown with the long bars;
for Galactic clusters, with the short bars.
The ages of the Worthey's (1994) models are given in gigayears.
The metallicities for the Worthey's models are
-2.00, -1.50, -1.00, -0.50, -0.22, 0.00, +0.25, +0.50,
if one takes the signs from left to the right.}
\end{figure*}

\setcounter{figure}{2}
\begin{figure*}
\hspace{-1.0cm}
\begin{tabular}{p{0.90\textwidth}}
\includegraphics[width=16.5cm,clip]{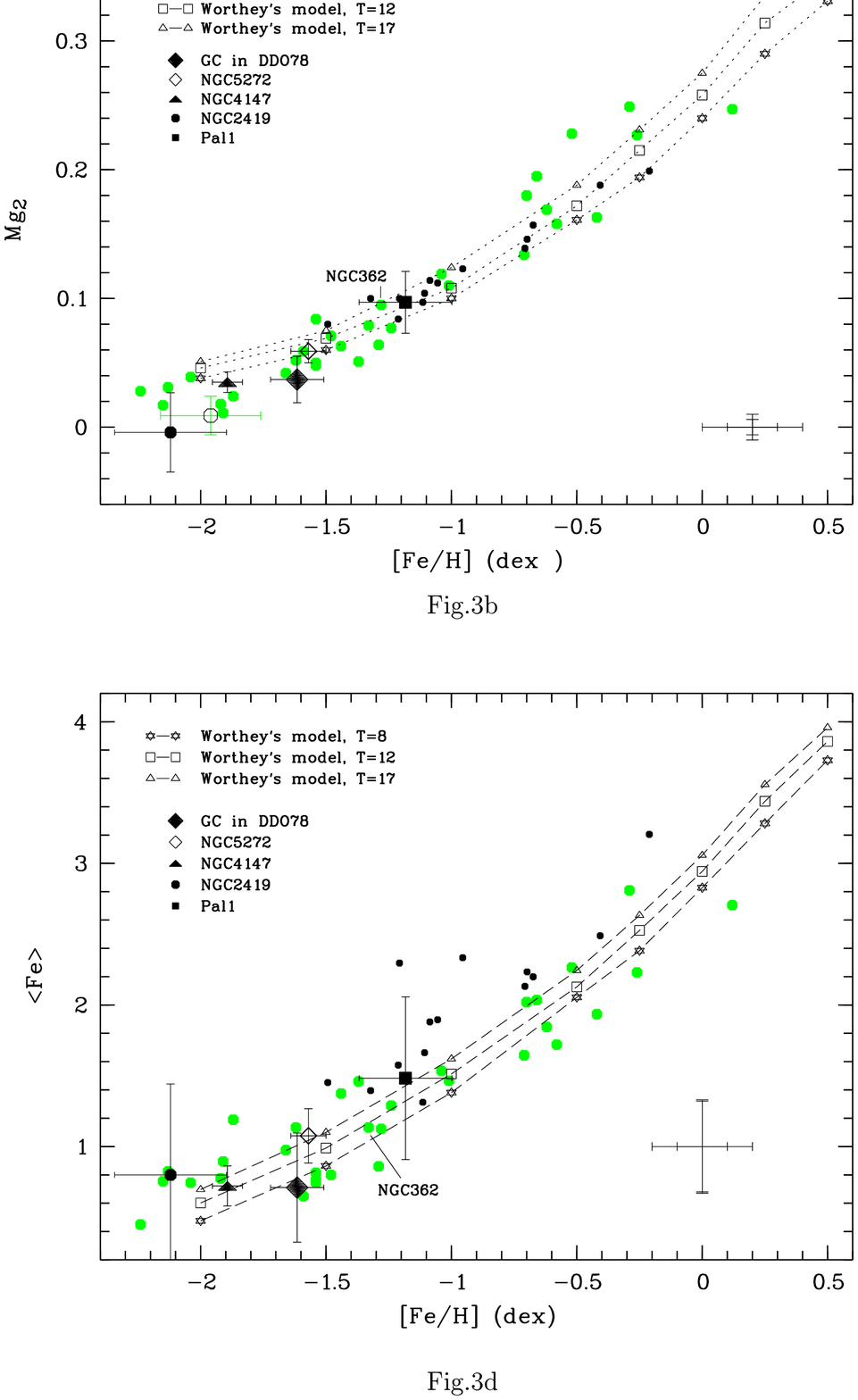} \\
\end{tabular}
\caption{ continued}
\end{figure*}

\setcounter{figure}{3}
\begin{figure*}
\hspace{-1.0cm}
\begin{tabular}{p{0.90\textwidth}}
\includegraphics[width=16.5cm,clip]{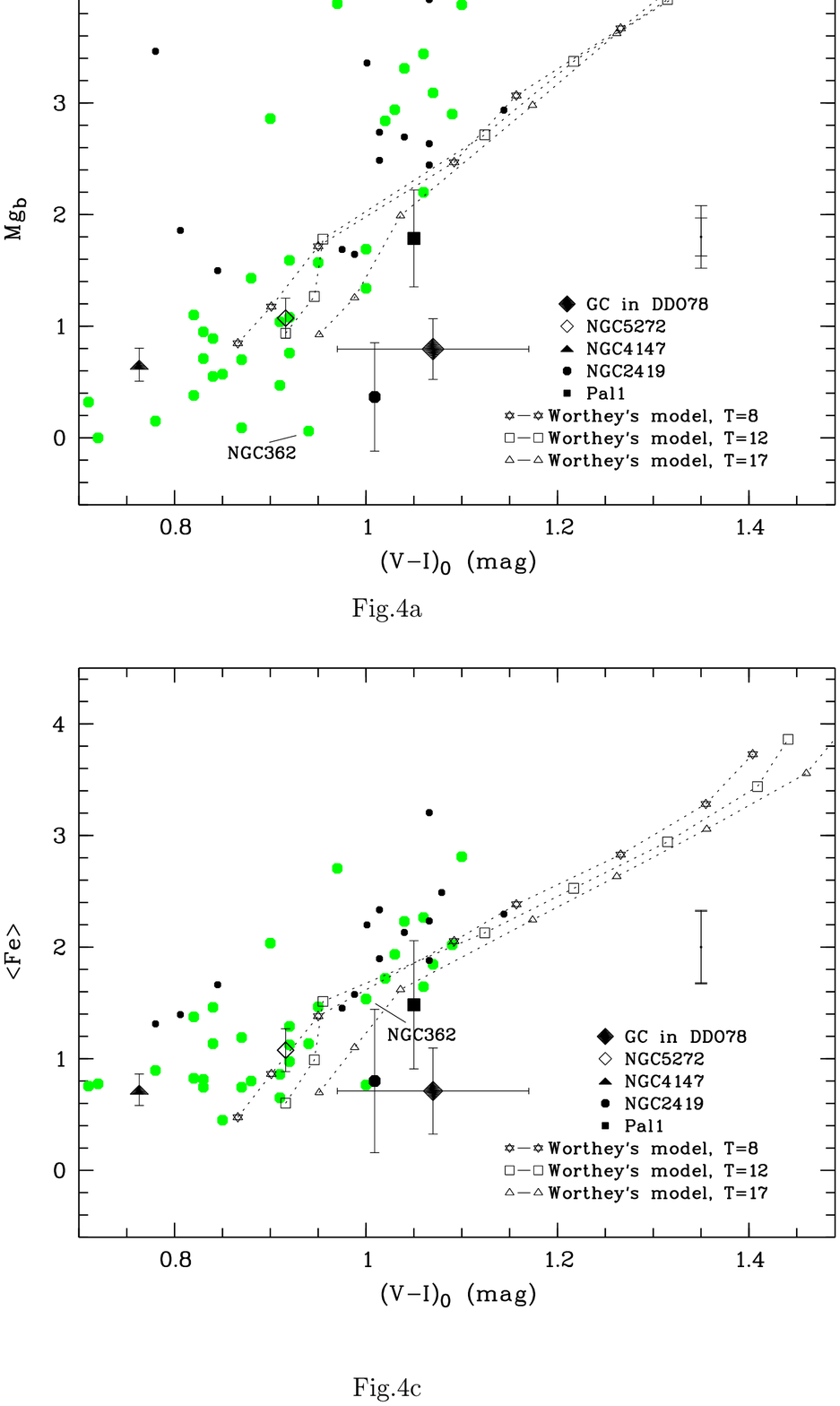} \\
\end{tabular}
\caption{Mgb index (Fig4a), Mg$_2$ (Fig4b), $\langle \mbox{Fe} \rangle$ (Fig4c)
and [Fe/H] (Fig4d) plotted as a function of $(V-I)_\mathrm{0}$ color.
The symbols are the same as in the Fig.3.
$(V-I)_\mathrm{0}$ color for GCs is taken from the catalog of Harris (1996).}
\end{figure*}

\setcounter{figure}{3}
\begin{figure*}
\hspace{-1.0cm}
\begin{tabular}{p{0.90\textwidth}}
\includegraphics[width=16.5cm,clip]{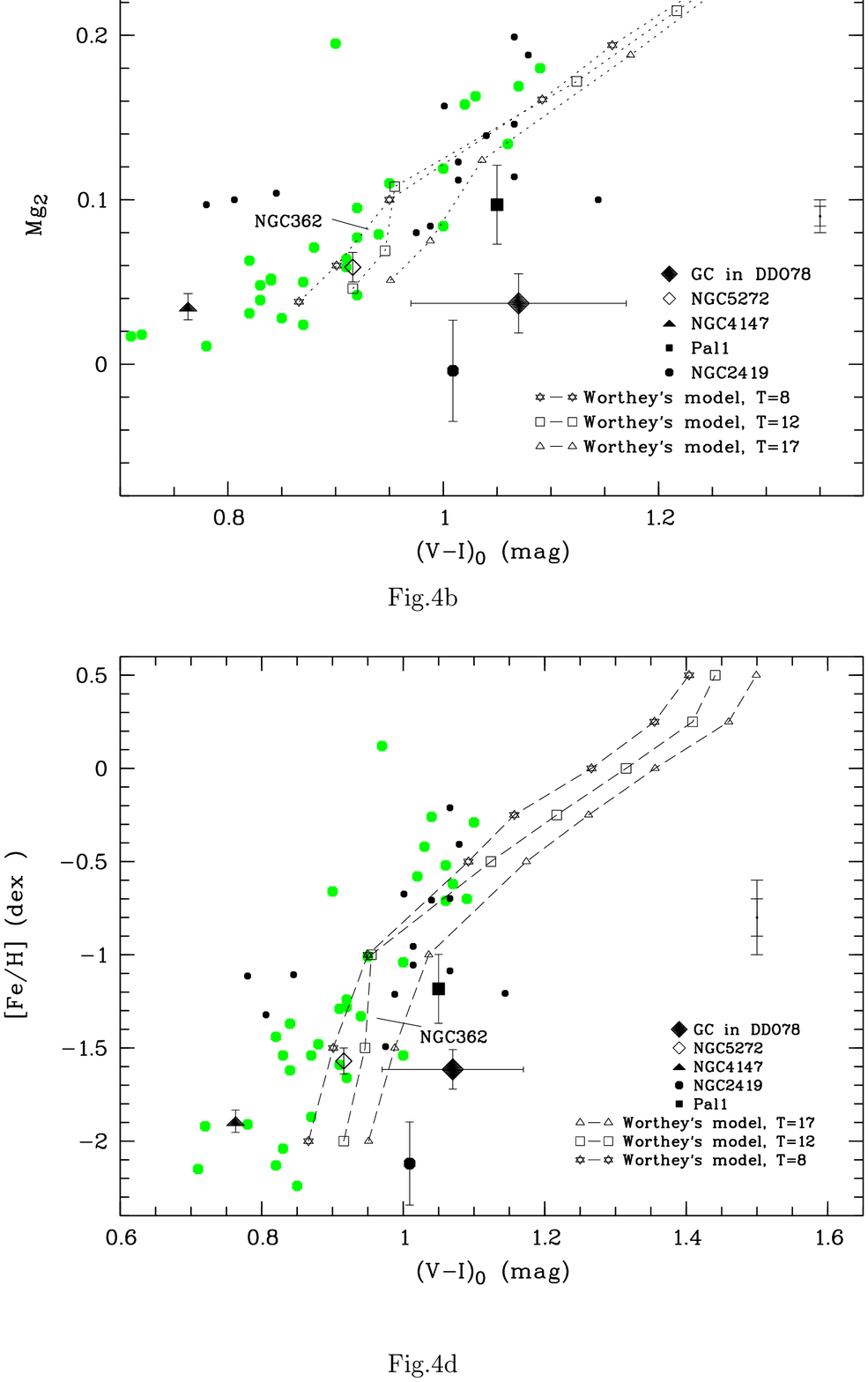} \\
\end{tabular}
\caption{ continued}
\end{figure*}

\setcounter{figure}{4}
\begin{figure*}
\hspace{-1.0cm}
\begin{tabular}{p{0.90\textwidth}}
\includegraphics[width=16.5cm,clip]{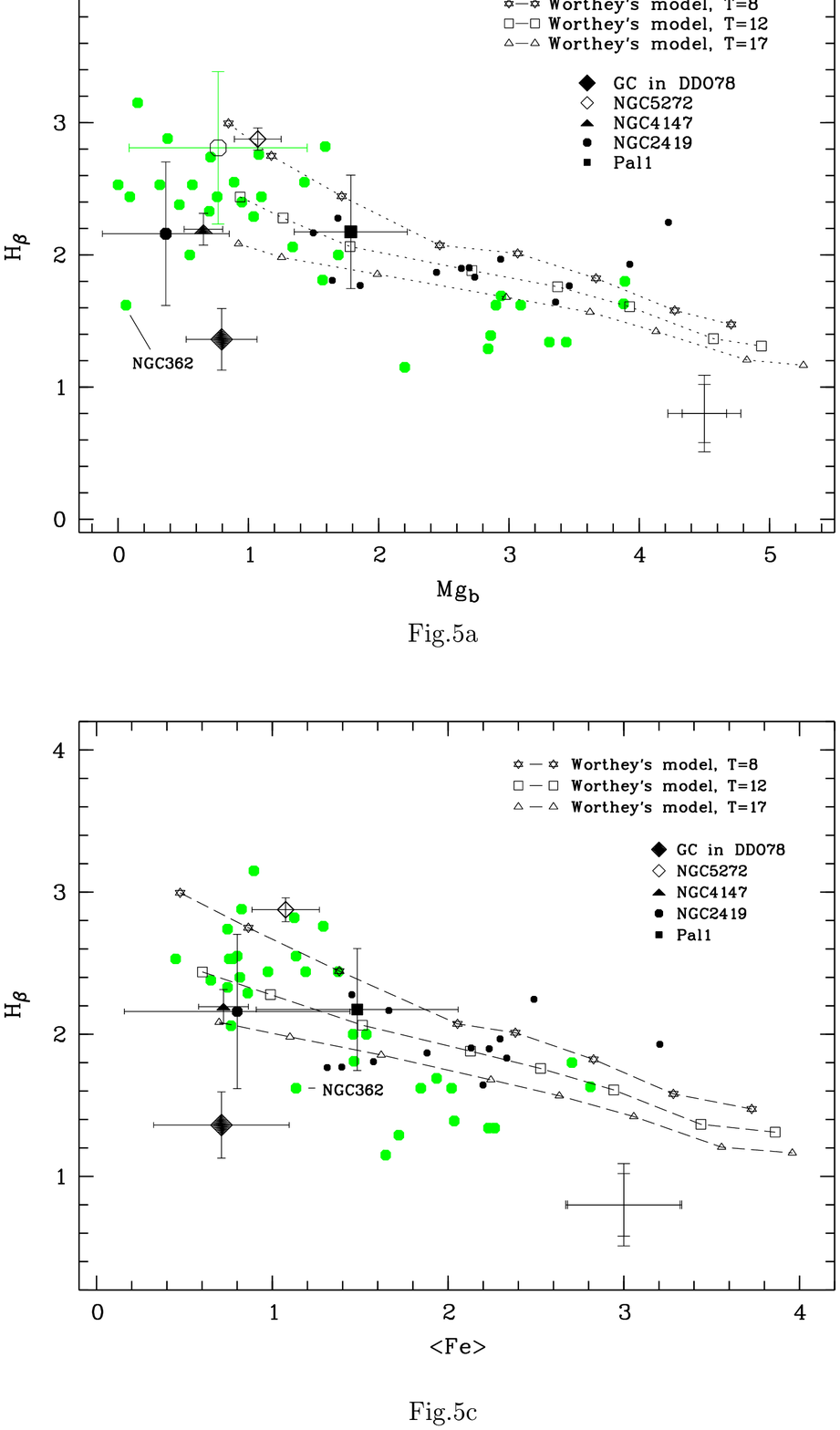}  \\
\end{tabular}
\caption{The age-sensitive index H$\beta $ plotted as a function of Mgb (Fig5a),
 Mg$_2$ (Fig5b), $\langle \mbox{Fe} \rangle$ (Fig5c) and [Fe/H] (Fig5d).
The symbols are the same as in the Fig.3.}
\end{figure*}
\setcounter{figure}{4}
\begin{figure*}
\hspace{-1.0cm}
\begin{tabular}{p{0.90\textwidth}}
\includegraphics[width=16.5cm,clip]{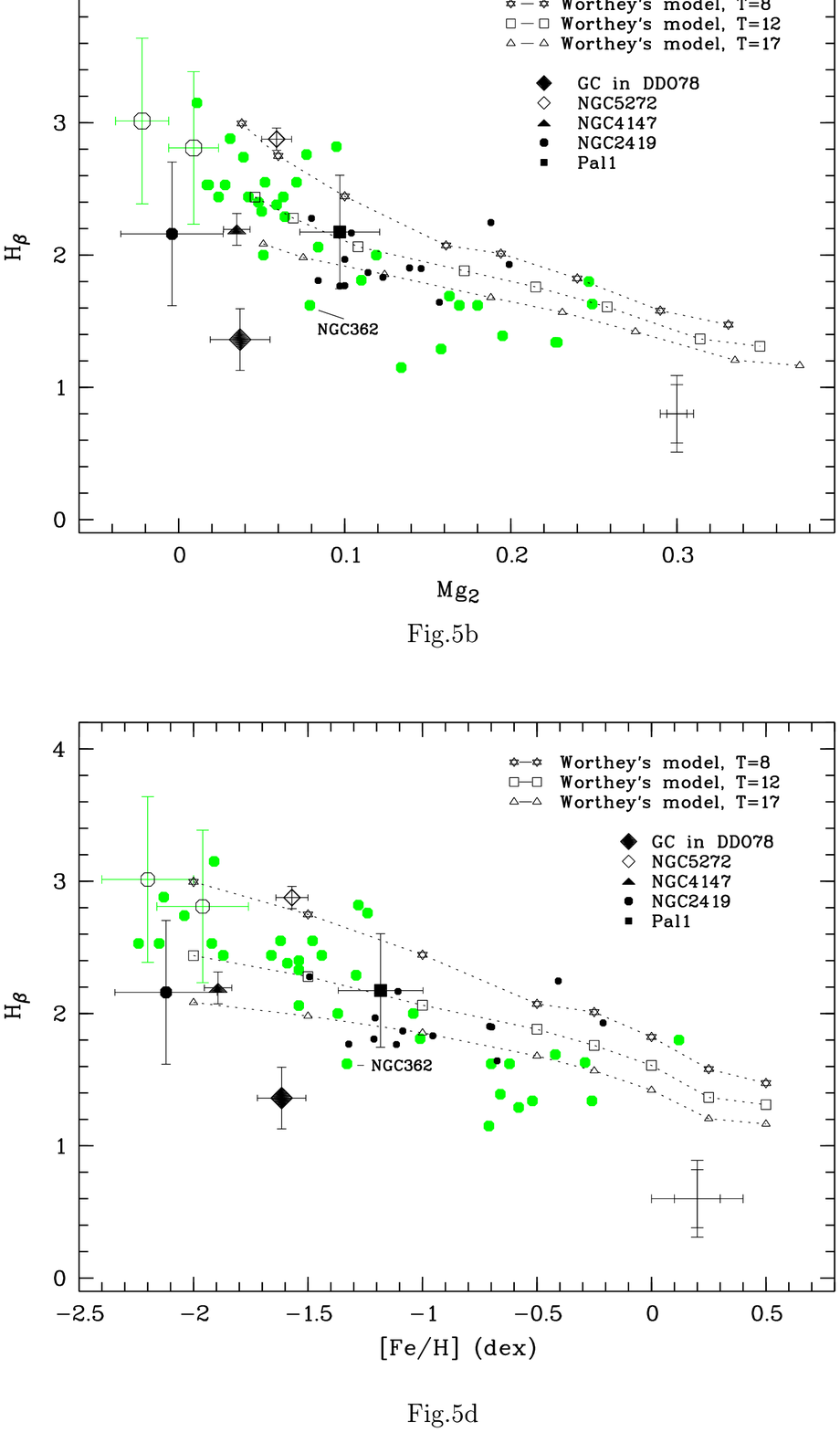}  \\
\end{tabular}
\caption{continued}
\end{figure*}

\end{document}